# Using multilevel modeling to evaluate science literacy and technology course of the Indonesian non-science students

Bayu Setiaji[a], Purwoko Haryadi Santoso[b,c]*, Khafidh Nur Aziz[a], Heri Retnawati[b], and Moh Khairudin[b]

[a]Department of Physics Education, Universitas Negeri Yogyakarta, Sleman 55281, Indonesia, [b]Graduate School of Educational Research and Evaluation, Universitas Negeri Yogyakarta, Sleman 55281, Indonesia; [c]Department of Physics Education, Universitas Sulawesi Barat, Majene 91413, Indonesia

* Correspondence: bayu.setiaji@uny.ac.id

Provide short biographical notes on all contributors here if the journal requires them.

# Using multilevel modeling to evaluate science literacy and technology course of the Indonesian non-science students

Science literacy is being fostered by science education community including the Indonesian education system. Science literacy and technology course has been designed and implemented to strengthen the national initiative empowering scientifically literate Indonesian society. This paper is intended to evaluate to what degree this course can be performed by non-science undergraduate students ($N = 160$) considering the nested structure of students' department and faculty setting. Multilevel modeling thus was employed to conduct the analysis dealing with this nature. The first level analysis involved students' performance and affective attribute measured using demonstrated science literacy assessment (SLA-D) and motivational beliefs (SLA-MB) respectively. The subsequent level analysis comprised demographic factors gathered from institutional record. Findings demonstrated that the impact of demographic factor to the students' performance on science literacy was not substantial. Different setting of students' department and faculty level drove the association between affective factor and learning process toward science literacy course substantially. Multilevel approach has controlled the equitable student assessment within the nature of students' data structure. This paper suggests many implications to open ideas regarding educational data analysis and to examine the effectiveness of science literacy course for the higher institution specifically for non-science majors.

Subject classification codes: include these here if the journal requires them

**Introduction**

Many research works have been attempted to develop scientific literacy by science education research community worldwide (Impey et al., 2011; Odden et al., 2021; Queiruga-Dios et al., 2020; Santoso et al., 2022). Survey reports established by the Programme for International Student Assessment (PISA) and the Trends in International Mathematics and Science Study (TIMSS) undoubtedly deliver major impact for the educational initiatives around the world including the Indonesian science education system (Cheema, 2017; Rachman et al., 2021; Ustun et al., 2022). Unsatisfying Indonesian students' performance as revealed by those survey program forces the government to revise their curriculum in addressing this challenge persuasively. Since 2020, the Indonesian Ministry of Education, Culture, Research, and Technology (KEMDIKBUDRISTEK) has piloted and disseminated curricular innovation in terms of 'MERDEKA BELAJAR' to shift the former national assessment program for the foci of scientific literacy education (Nurjati et al., 2022). Thereafter, the newer Indonesian assessment agenda is set to prepare secondary school students regarding science literacy domain as measured by the recent dimension of PISA and TIMSS survey (Salamah et al., 2022). Without neglecting the essence of content knowledge delivered by other disciplines, this breakthrough is planned to propagate the better students' performance than the past international science survey as administered by the reputable organizations like PISA and TIMSS. At the same time, we confidence this policy can be progressive and must invite the whole stakeholders and policy makers engaged in building the sustainability of the future Indonesian civilization including higher education as endeavored by this paper.

Higher education institution must be a groundbreaker to take part around the area of education and research expertise. Universitas Negeri Yogyakarta (UNY) is one of the Indonesian teachers' education institutions (TEIs) including science education of secondary school students for the whole nation. Recently, UNY has put more attention to strengthen the present national vision of fostering science literacy. At the same time, science literacy and technology course (MKU 6217) has been programmed since 2020 as compulsory according to the syllabus of the first-year undergraduate students across the department and faculty (BAKK UNY, 2019). This course is predominantly offered for non-science students to furnish their competencies in being the future Indonesian educators. Albeit they will fall into non-science carrier, they are prepared with several aspects of the nature of science encompassing the principles of scientific investigation. Essentially, they are anticipated to be set up as the imminent generation of society after they proceed from the undergraduate education. Meanwhile, they are still projected to be prospective teachers professionally on each field but are equipped to empower scientifically literate Indonesian society.

As previously highlighted, this course is crucial particularly for TEIs like UNY in educating the prospective teacher for the Indonesian community. UNY has just started the course implementation since 2020. This would be the appropriate moment to evaluate the implementation of the science literacy program for non-science students. Admittedly, effective science literacy instruction requires attention of evaluation during its implementation (Hobson, 2003). In this study, students' data have been harvested from four different non-science major in the 2022/2023 academic year, namely marketing, accounting, Javanese language, and dance education. Due to the distinct majors, students participated in this study evidently generate diverse background, prior knowledge, family resources, and other variables related to them. Then, students' performance throughout the course must be diverse and mixed. Consequently, it would be challenging to better understand the course impact to the students' science literacy. Within this context, our study needs a comprehensive approach in evaluating to what degree science literacy course can be performed by non-science students.

Relevant works can be cited to understand state of the art regarding how researchers have designed and evaluated science literacy instruction specifically in the context of higher education and more specifically for non-science major as focused by our study. Our research is unique and challenging with the argument of investigation of science literacy course for non-science college students that are rooted from diverse background institutionally. Plethora of methods have been approached in evaluating science literacy course in the interdisciplinary context of college students (Efthimiou & Llewellyn, 2004; Hamper & Meisel, 2020; Hobson, 2008; Impey, 2013; Parkinson & Adendorff, 2004; Ross et al., 2013; Sjöström & Eilks, 2018; Surpless et al., 2018). From the mentioned literature above, we can simply summarize that few researchers consider the nested impact emerged from the nature of students' data. In fact, students' data is mostly situated within the hierarchical of college administrative area. Thus, clustered context must be facing by our students and has potential influence for them. Underestimating this nature will contribute to the bias of students' performance evaluation that should be carefully avoided.

Broadly speaking, students' performance is one proxy of educational process that is commonly reported by education scholars. Many analytical approaches have been proposed to better measure, examine, and evaluate students' performance during the learning processes (Ding, 2019). Dealing with the issue of hierarchical data above, multilevel modeling, rooted from regression analysis, is then proposed to further consider the nested structure of students' learning (Finch et al., 2016). In this study, using multilevel modeling approach will enrich our understanding toward students' performance on science literacy course and the corresponding association with some potential factors such as affective attribute and demographic variables as frequently reported by scholars using PISA and TIMSS data (Ersan & Rodriguez, 2020; Mohammadpour et al., 2015; Ustun et al., 2022; You et al., 2021).

Utilizing multilevel modeling for data analysis requires one to take clustered level of data endorsed by data points into consideration. In the context of this study, our first level of data points are students. They can generate educational data and associated factors that will bring potential influences on the students' learning. Adjacent to the cognitive aspect, students can possess affective attribute during their learning process. Also, this embodies the first level of our data since it is generated from students. Afterward, departmental factor would contribute to the subsequent hierarchy of our students' data. It can be understood based on demographic variables recorded by institutional information system. Students are managed in certain department and faculty. Demographics variables created by the nested structure of students' department and faculty can correlate with the potential difference of students' performance (Kanim & Cid, 2020; Salehi et al., 2019; Simmons & Heckler, 2020). Therefore, multilevel modeling must be approached to make the analysis more comprehensive.

Prior works have extensively documented the association of students' performance toward affective attribute (Bellová et al., 2021; Fives et al., 2014; Fortus et al., 2022; Rudolph, 2020) and demographic variable (Ustun et al., 2022; You et al., 2021) within scientific literacy learning. They have argued that affective result can substantially influence students' scientific literacy. Nevertheless, several researchers' approach distinct construction of the affective measure. Thus, the implication of their results to the other contexts needs to be further examined. For instance, Fortus (2022) describes the definition of affective attribute into four constructs, namely interest, attitude, self-efficacy or self-concept, and motivation. Other idea has been proposed by Fives et al. (2014) that have developed the motivational belief measurement for their science literacy assessment (SLA).

Furthermore, demographic variables may contribute to the variance of the scientific literacy assessment as reported by Ustun et al. (2022) and You et al. (2021) in the context of PISA data. They discover that scientific literacy can be substantially influenced by demographics variables such as economic/ social/ cultural status (ESCS). Nevertheless, several scholars have shown that the demographic effects are mixed and still inconclusive (Simmons & Heckler, 2020). There must be other more important factors to open the investigation more comprehensive.

To complete the missing area of the prior works, this study is framed as an evaluation attempt toward the recent implementation of science literacy and technology course designed by UNY curriculum. The nested data structure of students' performance, affective attribute, and demographic variable are measured and analyzed using multilevel data modeling technique within three level of data (student, department, faculty). To guide this study, two research questions are proposed in the following.

RQ1. To what degree does the mean difference of students' performance and affective attribute on science literacy course vary within department level, faculty level, and demographic variable?

RQ2. To what degree does the department level and the faculty level vary the dependence between students' performance on science literacy course, affective attribute, and demographic variable?

Investigating the influential factors toward students' learning would inform scholars, educators, and practitioners in designing the equitable science learning. In this study, the recent implementation of science literacy and technology course has been evaluated based on data harvested from 2022/2023 UNY academic year. The analysis is conducted based on multilevel modeling result of the dependence between students' performance, affective attribute, and demographic variable clustered from the nature of college students' department and faculty setting. Evidence reported by this paper can be helpful for opening discussion room concerning educational data analysis dealing with the complex structure of students' data that must be warranted.

**Method**

What this paper aims to do is to evaluate the Indonesian non-science students' performance on science literacy course associating with affective attribute and demographic variables hierarchically gathered from four distinct departments and two different faculties involved in the study. For this purpose, the details of the research method are presented in the following sections.

*Context*

As briefly introduced in the preceding section, the current study was circumstanced within science literacy and technology course (MKU 6217) administered by Universitas Negeri Yogyakarta (UNY) during the first term of 2022/2023 academic year. This course was a compulsory program for non-science major. Students were taught concerning nature of science, scientific method, development of mankind civilization, universe as a system, science technology and implementation, and culminated with the topic regarding big ideas in science. The syllabus of science literacy course was designed according to the university model and the aforementioned national movement.

Due to the ongoing transition of pandemic situation, distance learning was still one of the learning modes. Moodle based learning management system (LMS) developed by UNY prior to the pandemic, 'BESMART' (Priyambodo, 2016; Surjono et al., 2017), was utilized in delivering, managing, and administering this course to the whole students. Figure 1 depicts the learning dashboard provided by BESMART for science literacy course.

>>> Figure 1 <<<

The first author (B.S.) and the third author (K.N.A.) were assigned as the lecturer of this course across four non-science undergraduate departments from two faculties. The participant of the study was determined in a voluntary basis without neglecting the representativeness of non-science students enrolled at UNY. They were recruited from department of marketing (*n* = 38), accounting (*n* = 39), Javanese language (*n* = 51), and dance education (*n* = 50). Department of marketing and accounting education were registered under the administration of the 'Faculty of Economics' and two remaining study programs were administered under the 'Faculty of Language and Arts'. In total, we counted that there were 178 students had participated in this study. Those students which were recorded as missing some data points either belong to one or some variables had been scheduled for the second survey session. Nevertheless, eighteen students should be deleted for the further data analysis due to their lack of awareness to our provided second chance.

*Measures*

The first level of students' data was collected using the in-class assessment including students' performance on science literacy course, midterm exam, and their affective attribute. Students' performance then would be the dependent variable. At the same level, midterm exam and affective attribute were simultaneously measured. It would be the predictors that might influence students' performance. Subsequently, each department and faculty tend to share the parallel characteristics that correspond to the demographic variables emerged from the students' data. In this study, the demographic data were compiled from institutional record. The department was the second level and the faculty was the third level of data.

A measure of scientific literacy assessment (SLA) disseminated by Fives et al. (2014) was employed to probe students' performance. SLA covered the 'demonstrated' (SLA-D) and the 'motivational belief' scientific literacy assessment (SLA-MB). Twenty-six multiple choice items of SLA-D was utilized as a proxy of students' performance of this course. SLA-MB was employed in our survey of affective attribute.

Formerly, the English version of SLA should be translated and adjusted to the context of Indonesian community. Content validity to two experts with the teaching and research experience on science literacy education for more than ten years had been conducted. They were the fourth and fifth author of this paper.

Five open-ended items were also administered by the midterm examination. We examined students in several aspects of science literacy and were also inspired from constructs explained by Fives et al. (2014) but in other form of open questions rather than dichotomous responses as measured by SLA-D above. Some evidence suggest that science literacy should be measured using the open-ended format rather than closed response (Miller, 1998). Moreover, grading rubric of the midterm exam was developed by the lecturer of this course (the first and third author of this paper). Face validity had been conducted to those experts parallel to the former validation step of the Indonesian version of SLA.

Theoretically, three constructs were demonstrated by SLA-MB underlying the measurement of affective attribute of science literacy. They were comprised of value of science (VOS), what can I do in science (DIS), and what I belief about science (BAS) (Fives et al. 2014). Twenty-five observed items of five-point Likert scale were distributed under these three factors. Using the students' response data ($N = 160$), we discovered plausible reliability as discovered by the Fives et al's result ($\alpha_{all} = 0.781$, $\alpha_{VOS} = 0.868$, $\alpha_{DIS} = 0.786$, $\alpha_{BIS} = 0.903$). Exploratory factor analysis (EFA) was also used to validate the alignment between the Fives et al's construct to the students' representation. We discovered three well-defined and unique components. The eigen values of SLA-MB were achieved more than one. Therefore, it was consistent with prior finding that three constructs of SLA-MB can be used independently of one another (Fives et al. 2014).

The demographic variables were collected from the institutional registrar record. In this study, ten demographic variables were harvested and examined in the further analysis. They were gender (1 = male, 2 = female), admission pathway (1 = 'SNMPTN' or national based college admission system via portfolio, 2 = 'SBMPTN' or national based college admission system via written test, 3 = 'SM Prestasi' or on-campus based college admission system via portfolio, 4 = 'SM Utul' or on-campus based admission system via written test), tuition funding (1 = subsidized, 2 = non-subsidized), scholarship holder (1 = yes, 2 = no), high school background (1 = science, 2 = non-science), residence (1 = North Sumatera, 2 = West Sumatera, 3 = Riau, 4 = Jambi, 5 = Bengkulu, 6 = Bangka Belitung, 7 = Jakarta, 8 = West Java, 9 = Central Java, 10 = Yogyakarta, 11 = East Java, 12 = East Kalimantan), father/ mother education (1 = master, 2 = bachelor, 3 = diploma, 4 = senior high school, 5 = junior high school, 6 = elementary school, 7 = uneducated), and father/ mother monthly income (1 = more than IDR4.000.000, 2 = IDR3.000.000–4.000.000, 3 = IDR2.000.000–3.000.000, 4 = IDR1.000.000–2.000.000, 5 = less than IDR1.000.000). Table 1 describes the summary of the context of participants on each class of each attribute.

>>> Table 1 <<<

We could discover that majority of our students were female (76%), admitted by university through on-campus admission via portfolio or 'SM Prestasi' (48%), non-subsidized funding (74%), no scholarship (84%), non-science high school background (67%), Central Java people (39%), live with senior high school graduated father (44%) and mother (43%), and economic status with father's and mother's monthly income between IDR 1 million – 2 million and less than IDR 1 million respectively.

In conclusion, we collected three variables from in-class assessment and ten variables extracted from university registrar record. For the requirement of the subsequent explanation, Table 2 summarizes the description of each variable employed in this study, including code (for the equation explanation below), measurement tool, and data type endorsed by each corresponding variable.

>>> Table 2 <<<

*Analysis*

Students' performance of this study as measured by SLA-D was scored using rubric of the correct option disseminated by Fives, et al. (2014). Of 26 multiple choice items, students' response was scored and transformed into 100 scale points. Then, before affective attribute as measured by SLA-MB (Aff) was summed up, students' response of BAS factor should be reversed due to the negative items. Midterm exam (MidTerm) was also graded using the same scale as determined in the students' performance.

Prior to the multilevel modeling, analysis of variance (ANOVA) was employed in RQ1 to test the mean difference of students' performance, midterm exam, and affective attribute among the class of each department, faculty, and demographic aspect. ANOVA was calculated to justify that department and faculty level can influence the variance of students' performance. As well, ANOVA determined potential factors that should be included in the equation of multilevel model (RQ2). Merely significant differences ($\alpha < 0.05$) among the class on each factor that would be included in the multilevel equations influencing the students' performance on science literacy as the target variable.

In RQ2, multilevel modeling approach was utilized dealing with the nested data structure of the students' department and faculty level (Finch et al., 2016). Multilevel modeling technique would be fit to process students' data clustered within four distinct departments and under two distinct faculties. Two-level modeling was first analyzed. It followed the three-level modeling involving the faculty level in the subsequent. Figure 2 shows the data structure with three levels in which $i$, $j$, and $k$ represent student, department, and faculty level respectively.

>>> Figure 2 <<<

The first step of analysis was building the null model or the baseline in which none of predictors was included in the equation. The null model was used as a baseline for model building and comparison. The null model was formulated as follow.

$$\text{SPerf}_{ij} = \gamma_{00} + U_{0j} + \varepsilon_{ij} \quad \text{(Two-level)} \quad (1)$$

$$\text{SPerf}_{ijk} = \delta_{000} + V_{00k} + U_{0jk} + \varepsilon_{ijk} \quad \text{(Three-level)} \quad (2)$$

As the reference, we adapted mathematical notation of multilevel modeling from Finch et al. (2016). Equation (1) refers to the two-level modeling and equation (2) accounts for the three-level modeling consecutively. $\text{SPerf}_{ij}$ refers to the students' performance of the $i$-th students under the $j$-th department, $\gamma_{00}$ refers to grand intercept mean of the $j$-th department, $U_{0j}$ accounts for the random effect of the $j$-th department, and $\varepsilon_{ij}$ term indicates the student-level random error that is not explained by the model. Equation (2) is a bit similar with the former. Yet, this is built for the higher level thus we find $k$ subscript indicating the level of students' faculty. As $\gamma_{00}$ above, $\delta_{000}$ is the grand intercept mean of the $j$-th department nested in the $k$-th faculty. Accordingly, $V_{00k}$ represents grand intercept mean of the $k$-th faculty. Eventually, $U_{0jk}$ refers to the random effect and $\varepsilon_{ijk}$ is the random error that cannot be explained by the model.

After that, the subsequent models were made based on the initial finding in RQ1 using ANOVA and complemented with intra class correlation (ICC) results based on the null model. Multilevel modeling was fit using `lmer` function provided by `lme4` package of R programming environment (Bates et al., 2015). The estimation approach from restricted maximum likelihood (REML) was selected since it has proven more accurate than maximum likelihood estimation (MLE) for estimating variance parameter

(Luo et al., 2021). Overall, those built models were then compared using Akaike information criterion (AIC) (Matuschek et al., 2017). As the rule of thumb, we should interpret the best model based on the lowest AIC discovered in our results.

**Results**

*Mean difference of students' performance and affective attribute among department level, faculty level, and demographic variable (RQ1)*

Initial information from RQ1 justified potential association within features discovered based on statistical analysis. ANOVA will test the mean difference of certain variable among cluster or group. In this study, the cluster must be department, faculty, and demographic variables as described specifically in Table 2. Then, ICC is a measure to what degree categorical variable such as demographic factor can correlate to certain variables for instance, students' performance, midterm exam, and affective attribute. Therefore, a significant result from ANOVA and a plausible ICC value of a predictor would be the consideration to take account those variables of the multilevel model. The results of RQ1 are given in Table 3 below.

>>> Table 3 <<<

First, we discovered significant mean difference of students' science literacy and midterm exam among the department and the faculty level ($p < 0.05$). Furthermore, there was plausible correlation between department and faculty level toward students' science literacy. Midterm exam also demonstrated a correlation with the department level, yet we discovered no correlation with the faculty level. Those results accommodated that students' performance on science literacy course would be empirically influenced by differences among the department and the faculty. Hence, one can justify equation (3) and (4) which add midterm exam (MidTerm) to the model.

$$\text{SPerf}_{ij} = \gamma_{00} + U_{0j} + \gamma_{10}\text{MidTerm}_{ij} + \varepsilon_{ij} \qquad \text{(Two-level)} \qquad (3)$$

$$\text{SPerf}_{ijk} = \delta_{000} + V_{00k} + U_{0jk} + \delta_{100}\text{MidTerm}_{ijk} + \varepsilon_{ijk} \quad \text{(Three-level)} \qquad (4)$$

Model 3 and 4 added the $\gamma_{10}$ and $\delta_{100}$ term. They enumerate the regression coefficient between its predictor and the outcome variable.

Second, we find no significant mean difference of students' affective attribute among department and faculty. Conversely, Fives et al. (2014) argued that there is strong correlation between affective measure and students' science literacy. However, Figure 2 could be consulted to support Fives et al. (2014) that depicts a scatter plot matrix of SPerf, Aff, and MidTerm Pearson correlation value. One can see that there is correlation between SPerf and Aff. To that reason, the affective attribute could not be neglected to understand its association toward students' science literacy. Thus, the third model added the affective attribute as formulated in the equation (5) and (6) below.

$$\text{SPerf}_{ij} = \gamma_{00} + U_{0j} + \gamma_{10}\text{MidTerm}_{ij} + \gamma_{20}\text{Aff}_{ij} + \varepsilon_{ij} \qquad \text{(Two-level)} \quad (5)$$

$$\text{SPerf}_{ijk} = \delta_{000} + V_{00k} + U_{0jk} + \delta_{100}\text{MidTerm}_{ijk} + \delta_{200}\text{Aff}_{ijk} + \varepsilon_{ijk} \quad \text{(Three-level)} \quad (6)$$

Clearly, $\gamma_{20}$ and $\delta_{200}$ accounted for the coefficient of the affective attributes toward the students' science literacy.

>>> Figure 2 <<<

Third, there is a significant mean difference ($p < 0.05$) of affective measure among the class of students' high school background and social status in terms of mother education. Surprisingly, there were no significant impact of other demographic variables such as gender, admission pathway, funding, scholarship holder, residence, father education, and parents' income. Hence, those non-significant demographic factors should be omitted. The next model should invite those results and we added two significant demographic variables (HS and MEdu) as explained by the equation (7) for the two level and (8) for the third level model as follow.

$$\text{SPerf}_{ij} = \gamma_{00} + U_{0j} + \gamma_{10}\text{MidTerm}_{ij} + \gamma_{20}\text{Aff}_{ij} +$$
$$\gamma_{30}\text{HS}_{ij} + \gamma_{40}\text{MEdu}_{ij} + \varepsilon_{ij} \quad \text{(Two-level)} \quad (7)$$
$$\text{SPerf}_{ijk} = \delta_{000} + V_{00k} + U_{0jk} + \delta_{100}\text{MidTerm}_{ijk} + \delta_{200}\text{Aff}_{ijk} +$$
$$\delta_{300}\text{HS}_{ij} + \delta_{400}\text{MEdu}_{ij} + \varepsilon_{ijk} \quad \text{(Three-level)} \quad (8)$$

where $\gamma_{30}$ and $\delta_{300}$ indicates the impact of different students' high school background toward their students' performance on science literacy course. Accordingly, $\gamma_{40}$ and $\delta_{400}$ corresponds to the dependence of students' performance on science literacy with the status of mother education.

***The impact of department and faculty level toward the dependence between students' performance, affective attribute, and demographic aspect (RQ2)***

Current presentation will demonstrate our multilevel modeling results after the fitting of those equations. The results are given in Table 4. To make the description easier to interpret for the readers, we commence the two-level modeling results (model 1, 3, 5, and 7) that will be followed up with three-level findings (model 2, 4, 6, and 8).

>>> Table 4 <<<

*Results for the Two-Level Models*

Model 1 in the second column of Table 4 was built to examine the department level association toward students' performance on science literacy. For the fixed effects, we discovered significant intercept ($\gamma_{00}$, $p < 0.05$) and the value within parentheses in Table 4 was the corresponding its standard error. This significant intercept indicated that department level could have the possibility to influence students' performance on science literacy that could be correlated with the random effects. Value reported by the section of random effects in Table 4 was the variance component and the corresponding standard deviation within parentheses. Intuitively, variance and standard deviation results could be interpreted as the extent to which the intercept ($U_{0j}$) varies by department level. Then, residual component reported by Table 4 was the $\varepsilon_{ij}$ term in equation (1).

Respectively, model 3 added the MidTerm variable into the equation. We also discovered significant results indicated that midterm examination could potentially influence the students' science literacy. In the model 3, we also discovered significant intercept ($\gamma_{00}, p < 0.05$) and coefficient of midterm exam ($\gamma_{10}, p < 0.05$). Comparing with the null model, the model 3 reported greater standard error on its intercept. Conversely, we found diminished pattern of random effects both by the variance and the standard components of the intercept ($U_{0j}$) and residual ($\varepsilon_{ij}$). This indicated that multilevel models more precisely estimated the standard errors for our parameters.

In terms of model 5, we added the affective attribute into the model. However, there was no significant intercept ($\gamma_{00}, p > 0.05$) yet it had the greater standard error than two previous models. The coefficient of midterm exam ($\gamma_{10}$) was still significant ($p < 0.05$) with similar result and decreased slightly with the model 3. Affective attribute ($\gamma_{20}, p < 0.05$) was discovered significantly influencing students' science literacy and this is consistent as reported by Fives et al. (2014). The diminishing pattern of intercept ($U_{0j}$) and residual ($\varepsilon_{ij}$) were also discovered as the former model.

Regarding model 7, non-significant coefficients were discovered both in the students' high school background ($\gamma_{30}, p > 0.05$) and the mother education variable ($\gamma_{40}, p > 0.05$). Those variables thus cannot be concluded as influential factor to predict students' science literacy. This result was inclusive differed from previous literature mentioned in the introduction above (Ersan & Rodriguez, 2020; Mohammadpour et al., 2015; Ustun et al., 2022; You et al., 2021). Intercept of the fixed effects ($\gamma_{00}, p > 0.05$) was non-significant as reported by model 5 of affective impact.

Finally, overall model has been reported in the last section of the Table 4. These results can be helpful to characterize and compare two-level models. Degree of freedom (*df*), number of groups (*N*), and observations are reported to characterize the models. Among four models, the lowest AIC was reported by model 5. Thus, model 5 was the best two-level model that fit the data analyzed in this study.

*Results for the Three-Level Models*

As practiced in the two-level modeling, null model was also built for the three-level model as the third column of Table 4. Significant intercept ($\delta_{00}$, $p < 0.05$) was also discovered with greater standard error than model 1 above. Therefore, incorporating the faculty level into the model capture the more accurate model to understand students' performance on science literacy. Then, variance and standard deviation of the intercept from the department level ($U_{0jk}$) were lower than the model 1 but with the same value of residual ($\varepsilon_{ijk}$). This remaining value can be shared to the added information varied by the faculty level ($V_{00k}$).

Midterm examination was significant factor predicting students' science literacy course ($\delta_{100}$, $p < 0.05$). This was consistent with the result reported by model 3 above. Obviously, greater standard error of significant intercept ($\delta_{000}$, $p < 0.05$) was also reported. For the random effects, we also discovered the diminished pattern of the intercept from department level ($U_{0jk}$) and the model residual ($\varepsilon_{ijk}$).

In terms of model 6, the affective attribute was combined into the model. Relevant with the model 5, it discovered the non-significant intercept ($\delta_{000}$, $p > 0.05$). Meanwhile, it still obtained the greater standard error than the model 4. The coefficient of midterm exam was significant ($\delta_{100}$, $p < 0.05$). Then, affective attribute was discovered significantly ($\delta_{200}$, $p < 0.05$) influencing students' science literacy and this is consistent as reported by model 5 in the two-level modeling result.

Regarding model 8, non-significant coefficients were discovered both in the high school variable ($\delta_{300}$, $p > 0.05$) and the mother education variable ($\delta_{400}$, $p > 0.05$). Hence, those variables were unable to interpret as influential factor to predict students' science literacy as reported by model 7 for two-level modeling results. Eventually, the lowest AIC among four three-level modeling results was reported by model 6 that merely considered students' level variable (midterm exam and affective attribute) into the model as performed by model 5 above. Therefore, model 6 was the best fit of the three-level model based on the data analyzed in this study.

**Discussion**

This study is proposed to answer two research questions. In RQ1, we investigate to what extent the mean difference of students' science literacy and associated factors can be varied by different department level, faculty level, and demographic variable. In RQ2, we study to what degree the discovered department and faculty effects in RQ1 contribute to investigate the association between affective attribute and demographic variables toward students' performance on science literacy course.

We can highlight four main findings for the answer of RQ1. First, students' department and faculty level have been evident as influential factor to make significant differences on their science literacy. This is consistent with prior study that has been mentioned in the introduction (Kanim & Cid, 2020; Salehi et al., 2019; Simmons & Heckler, 2020). Educational setting in which students are immersed during the learning process will construct their climate of learning. There are many factors incorporating the different context of students' learning process. It can be driven by complex factors in terms of teachers' quality, class size, peer motivation, physical facility, and another difficult factor to identify. Therefore, controlling the measurement of students' performance using the hierarchical data should be worth maintaining for the educators.

Second, the formative assessment during the learning process can be the best predictor in evaluating the peak of the students' performance on science literacy course. This is relevant with what has been emphasized Hobson (2003). He suggests that the effective science literacy instruction requires attention of evaluation during its implementation. Formative assessment including midterm examination is one of the assessment points in controlling students' eager during the learning process. Maintaining learning intention until the last part can predict students' success of learning science literacy in this study. This finding may be unsurprising since it can be the common pedagogical knowledge for most of the educators in many places.

Supplementing the influence of formative assessment directly, affective impact toward students' science literacy has been confirmed by our study significantly. This is consistent with prior works that have been mentioned in the introduction (Bellová et al., 2021; Fives et al., 2014; Fortus et al., 2022; Rudolph, 2020). It can be a recommendation for science literacy educators that quoting students attitudinal aspects in evaluating their performance must be quite imperative to create more equitable assessment. Controlling course evaluation regarding their associated affective factors is the principle of authentic assessment of science education (Nurjati et al., 2022, Salamah et al., 2022). Effective science literacy course should be more sensitive to elicit this factor for their class assessment criteria.

Of ten demographic variables collected in this study, we just discover two variables that make mean difference on students' science literacy. Surprisingly, gender that is mostly predicted to have great impact on science learning (Cheema, 2017, Ustun et al., 2022; You et al., 2021) does not give substantial factor to the variance of students' science literacy in this study. Meanwhile, high school preparation and mother education have been evident to contribute to influence the attainment of science literacy performed by our students. Further study must be warranted therefore this result can be described more extensively thus it can be more generalizable for the wider population.

Furthermore, we can discuss two main findings elaborated for the answer of RQ2 in the context of multilevel modeling results. First, we can summarize the best multilevel model and the pattern of the estimation results demonstrated by two level modeling above. Both two-modeling and three-modeling results agree that adding midterm exam and affective variable is the best model of students' science literacy assessment in this study. Adjacent to the cognitive aspect, affective attribute is the common procedure in doing authentic assessment as recommended by the literature of science literacy education (Nurjati et al., 2022, Salamah et al., 2022). This is relevant with the answer of RQ1 above. On the other hand, in model 5 we find no significant intercept of the random effects driven by department and faculty setting. This can be translated as controller of equitable assessment criteria for students' evaluation. Nevertheless, inviting affective measure to the consideration of assessment aspect can be carefully constructed since its latent factor that must be difficult to be measured.

Second, we discover the diminished pattern toward standard error of intercept after controlling the model using the multilevel analysis. This can be understood that variance source from department and faculty level precisely influence the dependent variable and we should consider for the better analysis. Precisely, it is evident that using hierarchical based analysis can capture the students' science literacy more accurate as recommended by the prior study (Ersan & Rodriguez, 2020; Mohammadpour et al., 2015. Hierarchical difference during the learning process is the consequences of college administrative area. Intuitively, this result can be a recommendation to the policy makers to ensure the equity in education for the whole students. Balancing the quality among the schools is admittedly able to make students feel more appreciated.

Nevertheless, we believe that this study may be driven by several sources of uncertainty caused by three factors. First, there could be errors and uncertainties introduced by our data collection methods. This study is framed as an evaluation attempt toward the implementation of science literacy course designed at UNY. Thus, the findings reported by this paper can be different in case of other universities owned by Indonesian education system. Second, affective measure is a latent construct. Many studies have reported diverse ideas to frame the definition of affective measure toward science literacy. In this study, our focus is not intended to characterize non-science majors differed from science majors. This can be a source of bias reported by paper written by science field. Thus, completing the current discussion with further study in comparison with science students must be highly recommended. The last but not the least, random intercept model is determined by our study due to the limitation to harvest higher level of predictors provided by the university information system. Engaging another multilevel model such random slope would be strongly suggested.

**Conclusion**

The Indonesian education system has been making a progressive policy to boost the more scientific literate community. Supporting this vision through higher education sector has been implemented by this paper via science literacy course designed for non-science Indonesian undergraduate students. Students are nested within department and faculty setting in the schema of college administrative system. Reporting the evaluation of science literacy course using multilevel model reveals some information regarding the effect of department and faculty setting toward students' performance throughout the course. In this study, most of the demographic aspects are unable to significantly influence the mean difference of students' science literacy. Instead, formative assessment and affective measure are two substantial factors that should be carefully considered in evaluating students' science literacy more equitable for the whole students. Evidence provided by this paper should be a recommendation to the higher institution in preparing the effective science literacy education for the prospective teachers of the future Indonesian science education.

Table 1. Description of participants based on the demographic attributes and the departmental distribution

| Attribute | Class | Faculty of Economics | | | | Faculty of Language of Arts | | | | Total | |
|---|---|---|---|---|---|---|---|---|---|---|---|
| | | Marketing | | Accounting | | Javanese | | Dance | | | |
| | | *n* | % | *n* | % | *n* | % | *n* | % | n | % |
| Gender | Male | 5 | .03 | 12 | .08 | 18 | .11 | 4 | .03 | 39 | .24 |
| | Female | 33 | .21 | 24 | .15 | 29 | .18 | 35 | .22 | 121 | .76 |
| Admission pathway | SNMPTN | 0 | 0 | 0 | 0 | 13 | .08 | 0 | 0 | 13 | .08 |
| | SBMPTN | 0 | 0 | 0 | 0 | 14 | .09 | 14 | .09 | 28 | .18 |
| | SM Prestasi | 35 | .22 | 24 | .15 | 5 | .03 | 12 | .08 | 76 | .48 |
| | SM Utul | 3 | .02 | 12 | .08 | 15 | .09 | 13 | .08 | 43 | .27 |
| Funding | Subsidized | 0 | 0 | 0 | 0 | 27 | .17 | 14 | .09 | 41 | .26 |
| | Non-subsidized | 38 | .24 | 36 | .23 | 20 | .13 | 25 | .16 | 119 | .74 |
| Scholarship | Yes | 8 | .05 | 1 | .01 | 6 | .04 | 10 | .06 | 25 | .16 |
| | No | 30 | .19 | 35 | .22 | 41 | .26 | 29 | .18 | 135 | .84 |
| High school Major | Science | 9 | .06 | 12 | .08 | 15 | .09 | 17 | .11 | 53 | .33 |
| | Non-science | 29 | .18 | 24 | .15 | 32 | .20 | 22 | .14 | 107 | .67 |
| Residence | North Sumatera | 0 | 0 | 2 | .01 | 0 | 0 | 0 | 0 | 2 | .01 |
| | West Sumatera | 0 | 0 | 0 | 0 | 0 | 0 | 1 | .01 | 1 | .01 |
| | Riau | 0 | 0 | 0 | 0 | 0 | 0 | 1 | .01 | 1 | .01 |
| | Jambi | 1 | .01 | 0 | 0 | 0 | 0 | 0 | 0 | 1 | .01 |
| | Bengkulu | 0 | 0 | 0 | 0 | 0 | 0 | 1 | .01 | 1 | .01 |
| | Jakarta | 0 | 0 | 2 | .01 | 1 | .01 | 0 | 0 | 3 | .02 |
| | West Java | 1 | .01 | 5 | .03 | 0 | 0 | 4 | .03 | 10 | .06 |
| | Central Java | 23 | .14 | 13 | .08 | 20 | .13 | 7 | .04 | 63 | .39 |
| | Yogyakarta | 10 | .06 | 14 | .09 | 17 | .11 | 14 | .09 | 55 | .34 |
| | East Java | 3 | .02 | 0 | 0 | 19 | .12 | 9 | .06 | 21 | .13 |
| | East Kalimantan | 0 | 0 | 0 | 0 | 0 | 0 | 1 | .01 | 1 | .01 |
| | West Nusa Tenggara | 0 | 0 | 0 | 0 | 0 | 0 | 1 | .01 | 1 | .01 |
| Father education | Master | 0 | 0 | 0 | 0 | 0 | 0 | 1 | .01 | 1 | .01 |
| | Bachelor | 11 | .07 | 4 | .03 | 11 | .07 | 10 | .06 | 36 | .23 |
| | Diploma | 3 | .02 | 0 | 0 | 2 | .01 | 1 | .01 | 6 | .04 |
| | Senior | 18 | .11 | 25 | .16 | 13 | .08 | 14 | .09 | 70 | .44 |
| | Junior | 0 | 0 | 3 | .02 | 10 | .06 | 7 | .04 | 20 | .13 |
| | Elementary | 5 | .03 | 4 | .03 | 9 | .06 | 5 | .03 | 23 | .14 |
| | No school | 1 | .01 | 0 | 0 | 2 | .01 | 1 | .01 | 4 | .03 |
| Mother education | Master | 2 | .01 | 0 | 0 | 0 | 0 | 0 | 0 | 2 | .01 |
| | Bachelor | 9 | .06 | 6 | .04 | 4 | .03 | 9 | .06 | 28 | .18 |
| | Diploma | 4 | .03 | 3 | .02 | 3 | .02 | 2 | .01 | 12 | .08 |

| Attribute | Class | Faculty of Economics | | | | Faculty of Language of Arts | | | | Total | |
|---|---|---|---|---|---|---|---|---|---|---|---|
| | | Marketing | | Accounting | | Javanese | | Dance | | | |
| | | *n* | % | *n* | % | *n* | % | *n* | % | n | % |
| | Senior | 16 | .10 | 13 | .08 | 21 | .13 | 18 | .11 | 68 | .43 |
| | Junior | 3 | .02 | 9 | .06 | 14 | .09 | 4 | .03 | 30 | .19 |
| | Elementary | 4 | .03 | 5 | .03 | 4 | .03 | 4 | .03 | 17 | .11 |
| | No school | 0 | 0 | 0 | 0 | 1 | .01 | 2 | .01 | 3 | .02 |
| Father income | > 4 mio | 7 | .04 | 7 | .04 | 2 | .01 | 4 | .03 | 20 | .13 |
| | 3–4 mio | 2 | .01 | 2 | .01 | 7 | .04 | 1 | .01 | 12 | .08 |
| | 2–3 mio | 9 | .06 | 6 | .04 | 5 | .03 | 6 | .04 | 26 | .16 |
| | 1–2 mio | 15 | .09 | 12 | .08 | 17 | .11 | 13 | .08 | 57 | .36 |
| | < 1 mio | 5 | .03 | 9 | .06 | 16 | 0.1 | 15 | .09 | 45 | .28 |
| Mother income | > 4 mio | 3 | .02 | 1 | .01 | 1 | .01 | 3 | .02 | 8 | .05 |
| | 3–4 mio | 7 | .04 | 1 | .01 | 4 | .03 | 2 | .01 | 14 | .09 |
| | 2–3 mio | 3 | .02 | 4 | .03 | 2 | .01 | 2 | .01 | 11 | .07 |
| | 1–2 mio | 6 | .04 | 4 | .03 | 4 | .03 | 6 | .04 | 20 | .13 |
| | < 1 mio | 19 | .12 | 26 | .16 | 36 | .23 | 26 | .16 | 107 | .67 |

Note: The colored cell is the majority class of each attribute

Table 2. Summary of investigated variables compiled by this study.

| No | Variable | Code | Source | Scale | Type |
|---|---|---|---|---|---|
| 1 | Students' performance | SPerf | 26 multiple choice items of SLA-D | 100-point | Continuous/ interval |
| 2 | Midterm exam | MidTerm | 5 open-ended items | 100-point | Continuous/ interval |
| 3 | Affective attribute | Aff | 25 questionnaire items of SLA-MB | 5-point Likert scale | Categorical/ ordinal |
| 4 | Gender | Gend | Institution registrar record | 2-point | Nominal |
| 5 | Admission pathway | Adm | | 4-point | Nominal |
| 6 | Funding | Fund | | 2-point | Nominal |
| 7 | Scholarship | Sch | | 2-point | Nominal |
| 8 | High school Major | HS | | 2-point | Nominal |
| 9 | Residence | Res | | 12-point | Nominal |
| 10 | Father education | FEdu | | 7-point | Nominal |
| 11 | Mother education | MEdu | | 7-point | Nominal |
| 12 | Father income | FInc | | 5-point | Nominal |
| 13 | Mother income | MInc | | 5-point | Nominal |

Table 3. ANOVA results of the mean difference of the students' performance (SPerf), affective attribute (Aff), and midterm exam (MidTerm) among the class on each students' department, faculty, and demographic factor

| Variable | SPerf | | | Aff | | | MidTerm | | |
|---|---|---|---|---|---|---|---|---|---|
| | *F* | *p* | ICC | *F* | *p* | ICC | *F* | *p* | ICC |
| Department | 11.21 | 0.000 | 0.209 | 0.162 | 0.922 | 0.000 | 6.902 | 0.000 | 0.129 |
| Faculty | 19.04 | 0.000 | 0.185 | 0.002 | 0.964 | 0.000 | 0.130 | 0.719 | 0.000 |
| Gender | 0.116 | 0.734 | 0.000 | 0.000 | 0.988 | 0.000 | 1.405 | 0.238 | 0.007 |
| Admission Pathway | 1.968 | 0.121 | 0.026 | 1.360 | 0.257 | 0.000 | 0.719 | 0.542 | 0.000 |
| Funding | 0.617 | 0.433 | 0.000 | 0.573 | 0.450 | 0.000 | 0.265 | 0.607 | 0.000 |
| Scholarship Holder | 0.600 | 0.440 | 0.000 | 0.008 | 0.929 | 0.000 | 0.010 | 0.920 | 0.000 |
| High School Major | 0.746 | 0.389 | 0.000 | 5.182 | 0.024 | 0.056 | 0.945 | 0.333 | 0.000 |
| Residence | 1.056 | 0.401 | 0.004 | 0.888 | 0.553 | 0.000 | 0.707 | 0.730 | 0.011 |
| Father Education | 0.867 | 0.521 | 0.000 | 1.329 | 0.247 | 0.030 | 0.613 | 0.720 | 0.000 |
| Mother Education | 1.162 | 0.330 | 0.000 | 2.245 | 0.042 | 0.055 | 0.228 | 0.967 | 0.000 |
| Father Income | 0.793 | 0.532 | 0.000 | 0.557 | 0.694 | 0.000 | 0.619 | 0.650 | 0.000 |
| Mother Income | 0.511 | 0.728 | 0.000 | 1.185 | 0.320 | 0.020 | 1.059 | 0.379 | 0.000 |

Note: The colored cells indicate that there is significant mean difference among the class of each students' department, faculty, and demographic aspect ($\alpha < 0.05$).

Table 4. Multilevel modeling results

| Parameter | Model 1 | Model 2 | Model 3 | Model 4 | Model 5 | Model 6 | Model 7 | Model 8 |
|---|---|---|---|---|---|---|---|---|
| **Fixed effects** | | | | | | | | |
| Intercept | 221.46*(9.545) | 221.54*(13.18) | 119.86*(26.47) | 116.86*(27.85) | 21.10(28.12) | 18.26(29.44) | -4.00(36.36) | -8.96(37.49) |
| Midterm exam | | | 1.27*(0.3132) | 1.31*(0.31) | 1.28*(0.28) | 1.32*(0.27) | 1.34*(0.27) | 1.39*(0.27) |
| Affective | | | | | 1.18*(0.18) | 1.19*(0.18) | 1.18*(0.19) | 1.18*(0.19) |
| High school | | | | | | | -5.72(5.20) | -5.55(5.19) |
| Mother Edu | | | | | | | 16.96(22.68) | 16.84(22.59) |
| **Random effects** | | | | | | | | |
| *Department-level* | | | | | | | | |
| Intercept | 332.7(18.24) | 168.1(12.96) | 248.9(15.78) | 63.58(7.974) | 240.7(15.51) | 52.47(7.24) | 238.5(15.44) | 35.06(5.92) |
| *Faculty-level* | | | | | | | | |
| Intercept | | 247.4(15.73) | | 274.31(16.56) | | 279.03(16.70) | | 299.21(17.30) |
| Residual | 1257.3(35.46) | 1257.2(35.46) | 1150.9(33.92) | 1150.75(33.92) | 905.2(30.09) | 905.06(30.08) | 890.0(29.83) | 889.90(29.83) |
| **Overall model** | | | | | | | | |
| *df* | 3 | 4 | 4 | 5 | 5 | 6 | 12 | 13 |
| *N* (Groups) | Dep (4) | Fac (2), Dep(4) | Dep (4) | Fac (2), Dep(4) | Dep (4) | Fac (2), Dep(4) | Dep (4) | Fac (2), Dep(4) |
| *N* (Observations) | 160 | 160 | 160 | 160 | 160 | 160 | 160 | 160 |
| AIC | 1610.6 | 1612.4 | 1596.68 | 1597.90 | 1560.00 | 1560.99 | 1564.06 | 1564.63 |

Note: * *p* < 0.05

Figure 1

BESMART e-learning dashboard of Science Literacy and Technology course

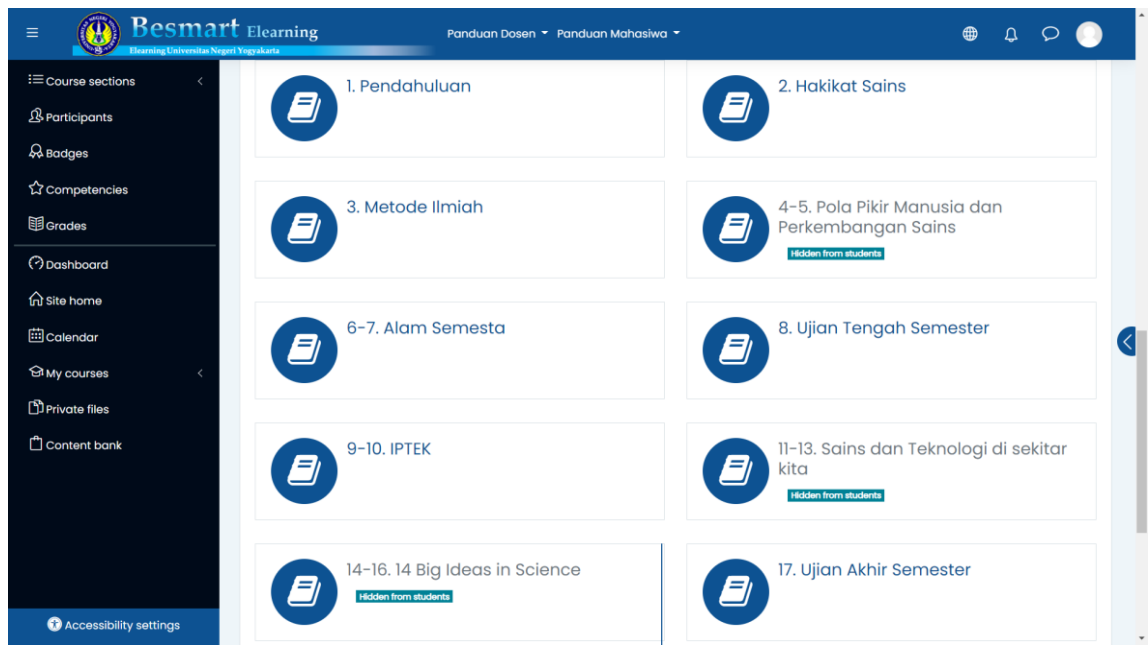

Figure 2. Network depicting data structure of student (Level 1), department (Level 2), and faculty (Level 3)

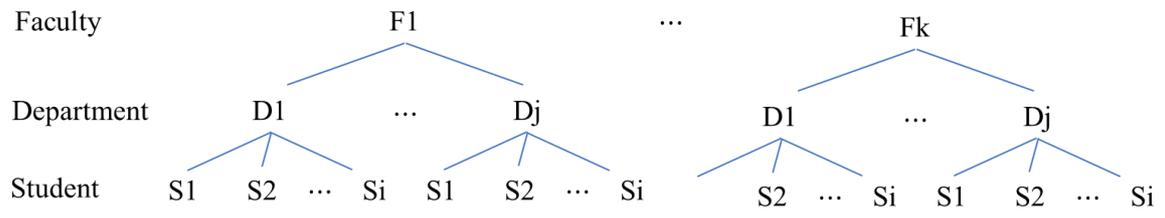

Figure 3. Scatter plot and Pearson correlation between SPerf, Aff, and MidTerm

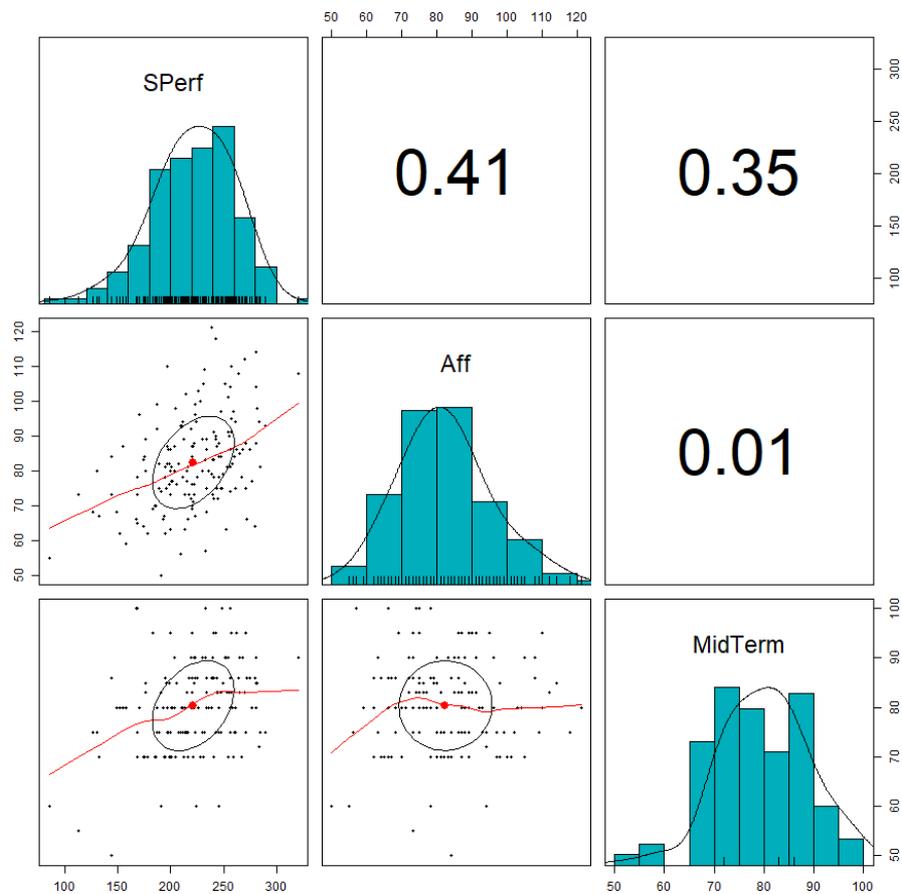

Note: Association between SPerf toward both Aff and MidTerm are described ($r = 0.41$ and $r = 0.35$ respectively). The independence within predictors (Aff and MidTerm) is also visualized ($r = 0.01$).